\documentstyle[pra,epsf,aps,twocolumn,epsf,psfig]{revtex}
\def\nablab{\mbox{\boldmath $\nabla$}}

\font\twlmbf=cmmib10 scaled1000

\newcommand{\gbf}[1]{\mbox{\twlmbf\symbol{#1}}}

\def\thetab{{\gbf{'022}}}

\def\deltab{{\gbf{'016}}}

\def\nablabs{\mbox{\boldmath $\scriptstyle\nabla$}}

\scriptsize

\def\aut#1{#1}
\begin{document}
\sloppy
\def\bpi{\begin{picture}}
\def\epi{\end{picture}}

\author{H.~Kleinert
                   \\
         Freie Universit\"at Berlin\\
          Institut f\"ur Theoretische Physik\\
          Arnimallee14, D-14195 Berlin
     }
\title{Vortex Origin of Tricritical Point
in Ginzburg-Landau Theory
}
\maketitle
\begin{abstract}
 Motivated
by recent experimental progress in the critical regime
of high-$T_c$ superconductors
we show how the
tricritical point in a superconductor
can  be derived from the Ginzburg-Landau theory
as a consequence of vortex fluctuations.
Our derivation
explains
why
usual
renormalization group arguments
always
produce a first-order transition,
in contrast to experimental evidence and
Monte Carlo simulations.
\end{abstract}

{}~\\
{\bf 1.}    The critical
regime of
old-fashioned superconductors
can be describes extremely well by the Ginzburg-Landau theory
\cite{Lan}
at the mean-field level \cite{SJT,TIN}. The reason is  the smallness of the
Ginzburg temperature interval
$ \Delta T_{\rm G}$
around the mean-field critical temperature $T_c^{\rm MF} $
where fluctuation become important \cite{Ginz}.
This made discussions of the order of the superconductive
phase transition started by
Halperin, Lubensky, and Ma in 1972 \cite{HLM}
rather academic.

The situation has changed with the
advent of
  modern high-$T_c$ superconductors
temperature superconductors.
In these the
Ginzburg
temperature interval is large enough to
study
field fluctuations and critical behavior.
Several experiments have found
a critical point
of the XY universality class
 \cite{TSB}.
In addition, there seems to be
recent evidence
for an additional critical behavior
associated with the so-called charged
fixed point \cite{TS}.
In view of future
experiments, it is important to understand
precisely the nature of critical fluctuations.

The Ginzburg-Landau theory \cite{Lan}
describes a superconductor
with the help of an energy density
\begin{eqnarray}
 {\cal H} (\psi,\nablab \psi,{\bf A},\nablab {\bf A})\!&=&\!\frac{1}{2}\!\left\{ \!
\left[(\nablab -ie{\bf A})\psi\right]^2
\!+\!\tau
|\psi|^2
+\frac{g}{2}
 |\psi|^4\right\}
   \nonumber\\&+&\!\frac{1}{2}\left(\nablab \times {{\bf A}}\right)^2,
\label{@ner}
\end{eqnarray}
where
$\psi(x), {\bf A}(x) $  are pair field and vector potential, respectively,
and $e$ is the charge of the Cooper pairs.
 The parameter $\tau \equiv T/T_c^{\rm MF}-1$
is a reduced temperature measuring the distance from the
characteristic temperature $ T_c^{\rm MF}$ at which the
$\tau $-term changes sign.
The theory needs gauge fixing, which is usually done
by setting
$\psi(x)= \rho(x) e^{\i \theta(x)}$, rewriting the
covariant derivative of $\psi$ as
\begin{equation}
D\psi=[i(\nablab \theta -e{\bf A}) \rho +\nablab  \rho ]e^{i\theta},
 \label{@tr}\end{equation}
and eliminating
the phase variable $ \theta(x) $ by a local
gauge transformation ${\bf A}\rightarrow {\bf A}+\nablab \theta/e$.
This brings
 $ {\cal H} (\psi,\nablab \psi,{\bf A},\nablab {\bf A})$  to the form
\begin{eqnarray}
 {\cal H}_1\!&=&\!\frac{1}{2}
(\nablab  \rho) ^2 \!+\!V( \rho )
\!+\!\frac{1}{2}\left(\nablab \times {{\bf A}}\right)^2
+\frac{ \rho ^2e^2}{2}{\bf A}^2,
\label{@ner1}
\end{eqnarray}
where $V( \rho )$ is the  potential of the $ \rho $-field:
\begin{equation}
V( \rho )=\frac{\tau}2
 \rho ^2
+\frac{g}{4}
  \rho ^4 .
\label{@POT}\end{equation}
The last term in (\ref{@ner1})
is the famous
Meissner-Higgs mass $m_A=
  \rho e
$ \cite{SJT,TIN}
of the vector potential
 ${\bf A}$,
whose analog
 in the gauge theory
of electroweak interactions to explain
why interactions are so much weaker than electromagnetic interactions.

At the mean-field level, the energy density
(\ref{@ner1})
describes a
second order phase transition.
It takes place if $\tau $
drops below zero
where
the pair field $\psi(x)$ acquires the nonzero
expectation value
$\left\langle \psi(x)\right\rangle = \rho _0= \sqrt{-\tau /g}$,
the order parameter
of the system. The $ \rho $-fluctuations around this value
have a {\em coherence length\/}
$\xi= 1/  \sqrt{-2\tau }$.
The
Meissner-Higgs mass term
in (\ref{@ner1}) gives rise to a finite {\em penetration
depth\/} of the magnetic field $ \lambda  = 1/m_A=
1/ \rho_0 e$.
The ratio
of the two length scales
$ \kappa\equiv
 \lambda / \sqrt{2}\xi$,
 which for historic reasons carries a factor $ \sqrt{2}$,
is the  Ginzburg
parameter
whose mean field value is
$  \kappa _{\rm MF}\equiv \sqrt{g/e^2} $.
Type I superconductors have small values
of $ \kappa $, type II superconductors have large values.
At the mean-field level,
the dividing line is
lies at
 $\kappa =1/ \sqrt{2}$.

In
 high-$T_c$ superconductors, field fluctuations become important.
These can be
taken into account by calculating the partition function
and field correlation functions
from the functional integral  %
\begin{equation}
Z=\int {\cal D}\hspace{-1pt} \rho \,\rho \, {\cal D}{\bf A} \,e^{ -\int d^3x\,  {\cal H}_1}
\label{@ZZ}\end{equation}
  (in natural units with $k_BT=1$).
So far, all approximations to $Z$
found since the initial work \cite{HLM}
have had
notorious
difficulties in accounting
for the order of the superconductive phase transition.
In
\cite{HLM},
simple renormalization group arguments
\cite{Kad}
 in $4- \epsilon $
dimensions
suggested that
the transition should be of first order.
The technical signal for this is the nonexistence
of an
infrared-stable fixed
point in the renormalization
group flow
of the coupling constants $e$ and $g$
as a function of the renormalization scale.
Due to the smallness
of the Ginzburg interval $ \Delta T_{\rm G}$, the
first order was never verified experimentally.
Since then, there has been much
work  \cite{MANY}
trying to find an infrared-stable fixed point  in higher loop orders or by
different resummations of the
divergent perturbations expansions.

\begin{figure}[tbhp]
\unitlength1mm
\begin{picture}(0,40)
\put(-3,0){\psfig{figure=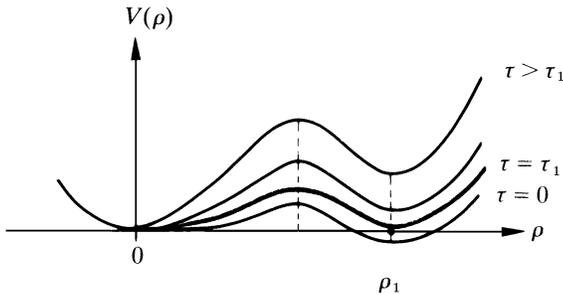,height=4cm}}
\end{picture}
\caption[]{Potential for the order parameter $ \rho $
with cubic term. A new minimum develops around $ \rho _1$
causing a first-order transition for $\tau =\tau _1$.}
\label{@p}\end{figure}

The simplest
argument
suggesting
a first-order nature of the transition arises
at the mean-field level of the pair field $ \rho $ as follows:
The
fluctuations of the vector potential
are Gaussian and can
 be
integrated out in (\ref{@ZZ}).
Assuming $ \rho $ to be smooth,
this may be done
in the Thomas-Fermi approximation \cite{PI},
leading to an additional cubic term
in the potential
(\ref{@POT}), which becomes
\begin{eqnarray}
V( \rho )=
\frac{\tau}2
 \rho ^2
+\frac{g}{4}
 \rho^4 -\frac{c}{3} \rho ^3,~~~~~c\equiv\frac{e^3}{2\pi}.\phantom{xx}
\label{@H2}\end{eqnarray}
The  cubic term
generates,
for $\tau <c^2/4g$,
a second minimum at \begin{equation}
 \tilde\rho _{0}=\frac{c}{2g}\left(1+\sqrt{1-\frac{4\tau g}{c^2}}\right).
\label{@newmin}\end{equation}
If $\tau $  decreases below
\begin{equation}
\tau _{\rm 1}=2c^2 /9g.
  \label{@tricr}\end{equation}
the new minimum lies lower than
the one at the origin
 (see Fig. \ref{@p}), so that the order parameter
jumps from zero to
\begin{equation}
 \rho _1=2c/3g
\label{@RHO1}\end{equation}
in a phase transition.
At this point,
the coherence length
of the $ \rho $-fluctuations
$\xi=1/\sqrt{\tau +3g \rho ^2-2c \rho }
$
has the finite value (the same as the fluctuations around $ \rho =0$)
\begin{equation}
\xi_1=   \frac{3}{c}  \sqrt{\frac{g}{2}} .
\label{@xi1}\end{equation}
The phase transition is therefore
of
first-order. 
\hspace{-2cm}

This conclusion is reliable
only if the
jump
 of $ \rho_0 $ is sufficiently large.
For small jumps,
the mean-field discussion of the energy
density (\ref{@H2})
cannot be trusted.
The place where the transition becomes second order
has, so far, never been explained satisfactorily
within the Ginzburg-Landau theory.
An explanation has been found
only
with the help of a dual disorder field theory
derived from
the Ginzburg-Landau theory
in Ref.~\cite{KLtr,GFCM}.
This theory is constructed in such a way
that its Feynman diagrams
are direct pictures
of the
vortex lines
of the superconductor.
The dual disorder field theory
shows
that there is a first-order transition
only if the
Ginzburg parameter $ \kappa\equiv  \lambda / \sqrt{2} \xi$
is smaller than  the tricritical value $ \kappa ^{\rm tric}\approx0.8/ \sqrt{2} $.
This point
is close
to the
mean-field value $ \kappa =1/ \sqrt{2} $
where the superconductor changes from type II to type I,
and the average short-range
 repulsion between vortex lines changes into an attraction, so that
the quartic term in the dual field theory becomes negative \cite{KLtr}.

In contrast to
the Ginzburg-Landau theory,
the vector potential  of the
 disorder field theory is massive from the outset,
so that its fluctuations do not
generate
a cubic term. Instead, they change
 the sign of the
{\em quartic\/} term, making
the transition first-order for $ \kappa < \kappa ^{\rm tric}$, while
leaving it second-order for $ \kappa > \kappa^{\rm tric}$.

The purpose of this note is to
show
how
the tricritical point
can be derived
from the
original Ginzburg-Landau theory
by a proper inclusion of fluctuation corrections.
The mistake
in the above argument
lies in the
neglect
of vortex fluctuations.
In fact, the
transformation  of the covariant derivative
$D\psi$
to the $ \rho$-$\theta$ expression
in Eq.~(\ref{@tr}) is false.
Since
$\theta({\bf x})$ and
$\theta({\bf x})+2\pi$
are physically indistinguishable ---
the complex
field
$\psi({\bf x})$ is the same for both ---
the correct substitution
is
\begin{equation}
D\psi=[i(\nablab \theta -2\pi \thetab^v-e{\bf A}) \rho +\nablab  \rho ]e^{i\theta},
 \label{@trc}\end{equation}
The
 cyclic nature of
the scalar field $\theta({\bf x})$
requires the presence of a vector field
$\thetab^v({\bf x})$
called
{\em vortex gauge field\/}.
This
field is a sum
of  $ \delta $-functions on Volterra surfaces
across which $\theta({\bf x}) $ has jumps by $2\pi$.
The boundary lines of the surfaces
are vortex lines. They are found
from the vortex gauge field
$\thetab^v({\bf x})$
by forming the
curl
\begin{equation}
\nablab\times  \thetab^v({\bf x})=  {\bf j}^v({\bf x}),
\label{@VD}\end{equation}
where
$
{\bf j}^v({\bf x})$ is
the {\em vortex density\/},  a sum over $ \delta $-functions
along the vortex lines $\deltab(L;{\bf x})\equiv \int_L d\bar {\bf x}\, \delta ({\bf x}-\bar{\bf x})$.
Vortex gauge transformations are deformations of the surfaces
at fixed boundary lines
which add to  $ \thetab^v({\bf x})$ pure gradients
of the form $\nablab
 \delta (V;{\bf x})$,
where
$ \delta (V;{\bf x})\equiv \int_V d^3\bar x\, \delta ({\bf x}-\bar {\bf x})$
 are $ \delta $-functions
on the volumes $V$ over which the surfaces  have swept.
The theory of these fields has been developed in the textbook
\cite{GFCM} and the Cambridge lectures \cite{CAM}.
Being a gauge field,   $ \thetab^v({\bf x})$
may be modified by a further
gradient of a smooth
function  to make it
purely transverse, $\nablab\cdot\thetab^v_T({\bf x}) =0$,
as indicated by the subscript $T$.
Since the vortex gauge field is
not a gradient, it cannot be absorbed
into the vector potential by a gauge transformation.
Hence
it survives
in the last term in
Eq.~(\ref{@ner1}), and
the
correct partition function is
\begin{eqnarray} \!\!\!
Z&\approx&\int {\cal D}\thetab^v_T \int
{\cal D}{ \rho } \rho \,
{\cal D}{\bf A}
\exp
\left[ -\frac{1}{2}(\nablab  \rho )^2  -\frac{\tau }2 \rho ^2-\frac{g}{4} \rho ^4
\right.\nonumber \\& &
 \left.~~~-\frac{1}{2}(\nablab\times {\bf A})^2
-\frac{\rho ^2e^2}{2}
({\bf A}-2\pi\thetab_T^v/e)^2\right]
.
\label{@XY1}\!\!\!\!\!\!\end{eqnarray}
The symbol
$
\int {\cal D}\thetab^v_T$ does not denote
an ordinary functional integral.
It is defined as a sum over all numbers
and shapes
of Volterra surfaces $S$
in
$
\thetab^v_T$, across which the phase jumps by $2\pi$ \cite{CAM}.

The important observation is  now that
the partial partition function  contained in (\ref{@XY1})
\begin{eqnarray}
Z_1[\rho]&\equiv& \int {\cal D}\thetab^v_T {\cal D}{\bf A}
 \exp\left\{
-\frac{1}{2}\int d^3x(\nablab\times {\bf A})^2
\right.\nonumber \\
&&~~~~~~~~~~~~~~~~~~~~~~~\left.
-
\frac{\rho ^2}{2}\int d^3x[e{\bf A}-2\pi\thetab_T^v]^2
\right\}
\label{@EFF}\end{eqnarray}
can give rise to a second-order transition
of the $XY$-model type if
the Ginzburg parameter
 $ \kappa $ is sufficiently large.
To see this we integrate
out the ${\bf A}$-field
and obtain
\begin{eqnarray}
&&Z_1[\rho]=
  \exp\left[ \int d^3x\frac{e^3 \rho ^3}{6\pi}\right]
\int {\cal D}\thetab^v_T
\label{@EFF2} \\
 &&\times
 \exp\left[
\frac{4\pi^2\rho ^2}{2}\int d^3x\,\left(
\frac{1}{2}\thetab_T^v\,{}^2
-
\thetab_T^v\,\frac{ \rho ^2e^2}{-\nablab^2+ \rho ^2e^2}
\thetab_T^v\right)\right]
.      \nonumber
\end{eqnarray}
The second integral
can be simplified to
\begin{eqnarray}
\frac{4\pi^2\rho ^2}{2}\int d^3x\,\left(
\thetab_T^v\,\frac{- \nablab^2}{-\nablab^2+ \rho ^2e^2} \thetab_T^v\right)  .
\label{@EFF2p}\end{eqnarray}
Integrating this  by parts, and replacing
 $\nabla_i\thetab_T^v\,\nabla_i \thetab_T^v$  by $
(\nablab\times  \thetab_T^v)^2=
{\bf j}^{v\,2} $,
the partition function
(\ref{@EFF2}) without
the prefactor takes the form
\begin{eqnarray}
Z_2[\rho]\!&=&\!\!
 \int \!{\cal D}\thetab^v_T
\exp\left[ -\frac{4\pi^2\rho ^2}{2}\int d^3x\,
\left(
{\bf j}^v
\frac{1}{-\nablab^2+ \rho ^2e^2}{\bf j}^v
 \right)
\right] . \nonumber \\
\label{@EFF3}\end{eqnarray}
This is the partition function
of the grand-canonical ensemble of closed fluctuating vortex lines.
The interaction between them has a finite range
equal to the penetration depth $ \lambda =1/ \rho e$.

It is well-known how to compute pair and magnetic fields
of the Ginzburg-Landau theory
for a single straight vortex line from the
extrema of  the energy density \cite{SJT}.
In an external magnetic field, there exist triangular and various
other regular
arrays of vortex lattices and various
phase transitions.
In the core of each vortex line,
the pair field $ \rho $ goes to zero over a distance $\xi$.
If we want to sum over grand-canonical ensemble
of fluctuating vortex lines of any shape
in the partition function (\ref{@XY1}),
the
space dependence of $ \rho $
causes complications.
These can be avoided by an approximation, in which
the system
is placed
on a simple-cubic lattice
of spacing $a=  \alpha\, \xi$,
with $ \alpha $ of the order of unity, and
a {\em fixed\/} value $ \rho = \tilde\rho _0$
given by
Eq.~(\ref{@newmin}).
Thus we replace the partial partition function
(\ref{@EFF3}) approximately by
 \begin{eqnarray}
Z_2[\tilde\rho_0] \!
&=&\!\!\sum_{\{{\bf l}; \nablabs\cdot{\bf l}=0\}}
\!\! \exp\left[ -
\frac{4\pi^2\tilde \rho_0 ^2a}{2} \sum_{\bf x}
{\bf l}({\bf x})
v_{ \tilde \rho_0 e}({\bf x}-{\bf x}')
{\bf l}({\bf x}')
\right] .  \nonumber \\
\label{@EFF4}\end{eqnarray}
The sum runs over the discrete
versions of the vortex density in (\ref{@VD}).
These are
integer-valued vectors
${\bf l}({\bf x})=
(
l_1({\bf x}),
l_2({\bf x}),
l_3({\bf x})) $ which satisfy
$\nablab\cdot {\bf l}({\bf x})=0$,
where
$\nablab$ denotes the lattice derivative.
This condition restricts the sum over all ${\bf l}({\bf x})$-configurations in
(\ref{@EFF4})
to all non-selfbacktracking
integer-valued  closed loops.  The function
\begin{eqnarray}
v_m({\bf x)}&=&\prod_{i=1}^3\int \frac{d^3(ak_i)}{(2\pi)^3}\frac{e^{i(k_1x_1+k_2x_2+k_3x_3)}}
{2\sum_{i=1}^3(1-\cos ak_i)+a^2m^2}\nonumber \\
&=&\int ds e^{-(6+m^2)s}
I_{x_1}(2s)
I_{x_2}(2s)
I_{x_3}(2s).
      \label{@YP}\end{eqnarray}
is the lattice
Yukawa potential \cite{GFCMLY}.

The lattice partition function (\ref{@EFF4})
is known to have a second-oder phase transition
in the universality class of the $XY$-model.
This can be seen by
a comparison with
the
Villain approximation
\cite{GFCMVIL}
to the $XY$ model,
whose partition function
is a lattice version of
\begin{eqnarray}
&&Z_V[\rho]=
\int {\cal D}\theta
\int {\cal D}\thetab^v_T
\label{@EFF7}
 \exp\left[ -\frac{b}{2}
\int d^3x\,\left(\nablab\theta
-\thetab_T^v\right)^2
\right]
.      \nonumber
\end{eqnarray}
After integrating out $\theta({\bf x})$, this becomes
\begin{eqnarray}
&&Z_V[\rho]={\rm Det}^{-1/2}(-\nablab^2)
\int {\cal D}\thetab^v_T
 \exp\left( -\frac{b}{2}
\int d^3x\,
\thetab_T^v\,{}^2\right),
\nonumber \\
\end{eqnarray}
and we can replace
 $\thetab_T^{v\,2}$  by $
\nablab\times  \thetab_T^v(-\nablab^2)^{-1}
(\nablab\times  \thetab_T^v)^2=
{\bf j}^v(-\nablab^2)^{-1}
{\bf j}^v $.
By taking this expression to a simple-cubic lattice
we obtain the partition function
(\ref{@EFF4}),
but with
$\tilde  \rho _0^2 a$ replaced by $ \beta_V\equiv ba$,
and  the Yukawa potential $v_{ \tilde \rho_0 e}({\bf x})$
replaced by
the Coulomb potential
$v_{0}({\bf x})$.

The partition function
(\ref{@EFF4}) has the same transition
at roughly the same place as
its local approximation
 \begin{eqnarray}
Z_2[\tilde\rho_0]
&\approx &\sum_{\{{\bf l}; \nablabs\cdot{\bf l}=0\}}
 \exp\left[ -
\frac{4\pi^2\tilde \rho_0 ^2a}{2}v_{ \tilde \rho_0 e}({\bf 0}) \sum_{\bf x}
{\bf l}^2({\bf x})\right] .
\label{@EFF5}\end{eqnarray}
A similar approximation holds for the Villain model
with
$v_{0}({\bf x})$ instead of
$v_{ \tilde \rho_0 e}({\bf x})$
and
$\tilde  \rho _0^2 a$ replaced by $ \beta_V\equiv ba$.

The Villain model is known to undergo a second-order phase transition
of the $XY$-model type at
$ \beta_ V= r/3$ with $r\approx1$,
where
the vortex lines become infinitely long
\cite{GFCM1236}.
 Thus we conclude that also the partition function
(\ref{@EFF5}) has a second-order phase transition
of the $XY$-model type at
$\tilde \rho ^2  v_{ \tilde \rho_0 e}({\bf 0})a\approx v_{0}({\bf 0})/3$.
The potential (\ref{@YP})
at the origin has the hopping expansion \cite{GFCMv}
\begin{equation}\!\!\!\!\!\!
v_m({\bf 0})=\!\sum_{n=0,2,4}
 \frac{H_n}{(a^2m^2+6)^{n+1}},\,H_0=1,H_2=6,\dots\,. \!\!\!\!\!\!\!
\label{@}\end{equation}
To lowest order, this yields the ratio
$ v_m({\bf 0})/v_0({\bf 0})\equiv 1/(m^2/6+1)$.  A more accurate
 fit to the
ratio
$
 v_m({\bf 0})/v_0({\bf 0})$ which is good up to $m^2\approx10$
(thus comprising all interesting $ \kappa $-values
since $m^2$ is of the order of $3/ \kappa ^2$)  is
$1/( \sigma \, m^2/6+1)$ with $ \sigma \approx1.38$.
Hence the
transition takes place at
\begin{equation}
\frac{ \tilde \rho_0^2a}{( \sigma \, a^2 \tilde\rho _0^2e^2/6+1)}
\approx\frac{r}3~{\rm or}~~ \tilde\rho _0\approx \frac{1}{ \sqrt{3a} }
\sqrt{\frac{r}{1- \sigma rae^2/18} } .
\label{@rho0}\end{equation}
This transition will persist until $ \tilde\rho _0$
reaches the value
$
 \rho _1=2c/3g$ of Eq.~(\ref{@RHO1}).
From there on, the transition
will no longer be of the $XY$-model type but
occur discontinuously as a first-order transition.

Replacing in (\ref{@rho0})
$a$ by $ \alpha \xi_1$ of Eq.~(\ref{@xi1}), and
 $\tilde \rho _0$ by $
\rho _{1}$, we find the
  equation for the mean-field Ginzburg parameter
$ \kappa_{\rm MF} = \sqrt{ g/e^2\/}$:
\begin{equation}
   \kappa ^{3}_{\rm MF}+  \alpha ^2 \sigma \frac{{ \kappa_{\rm MF} } }3-
\frac{ \sqrt{2} \alpha  }{\pi r}=0.
\label{@}\end{equation}
 Inserting $ \sigma \approx 1.38$ and
choosing $ \alpha \approx r\approx 1$, the solution of this yields the tricritical value
\begin{equation}
\kappa_{\rm MF} ^{\rm tric}\approx0.81/ \sqrt{2}.
\label{@}\end{equation}
In spite of the roughness
of the approximations,
this result is very close
to the value
$0.8/ \sqrt{2}$ derived from the dual theory in \cite{KLtr}.
The approximation has three uncertainties.
First, the identification
of the effective lattice spacing
$a= \alpha \xi$ with $ \alpha \approx1$;
second the
associated neglect of
the ${\bf x}$-dependence of $\rho $ and its
fluctuations,
and third
the localization of the
critical point of the $XY$-model type transition
in Eq.~(\ref{@rho0}) setting $q\approx1$.
By modifying slightly the parameters
$q$ and $r$ we can, of course,
obtain complete agreement with the previous result..

Our goal has been achieved:
We have shown the existence
of a
tricritical point in a superconductor
directly
within the fluctuating Ginzburg-Landau theory,
by taking the vortex fluctuations into account.
For $ \kappa>0.76/ \sqrt{2}  $ these
 give rise
to an $XY$-model type second-order transition
before the cubic term  becomes relevant.
The cubic term causes
 a discontinuous transition
only for $ \kappa<0.76/ \sqrt{2}  $.

{}~\\
Acknowledgment: \\
The author is
grateful to
F. Nogueira
for many
 valuable discussions.

\end{document}